\documentclass{mn2e}
\usepackage{psfig}
\def\etal{et~al. }
\def\la{\mathrel{\hbox{\rlap{\hbox{\lower4pt\hbox{$\sim$}}}{\raise2pt\hbox{$<$}}}}}
\def\ga{\mathrel{\hbox{\rlap{\hbox{\lower4pt\hbox{$\sim$}}}{\raise2pt\hbox{$>$}}}}}

\begin{document}

\title[Photometric variability of GD\,356]
{Photometric variability of the unique magnetic white dwarf
  GD\,356}

\author[C.\,S. Brinkworth \etal]
{C.\,S. Brinkworth$^1$, M.\,R. Burleigh$^2$, G.\,A. Wynn$^2$,
  T.\,R. Marsh$^{1,3}$\\
\\
$^1$ Department of Physics and Astronomy, University of Southampton,
  Highfield, Southampton, UK. \\
$^2$ Department of Physics and Astronomy, University of Leicester,
  Leicester, LE1 7RH, UK. \\
$^3$ Department of Physics, The University of Warwick, Coventry, CV4
7AL, UK. \\}
%\date{December 11th 2003}

\maketitle

\begin{abstract}

GD\,356 is a magnetic white dwarf (B $=13$MG) that uniquely displays
weak resolved Zeeman triplets of H$_\alpha$ and H$_\beta$ in
emission. As such, GD\,356 may be the only known white dwarf with some
kind of chromosphere, although accretion from the interstellar medium
or more exotic mechanisms cannot be ruled out.
Here, we report the detection of low amplitude ($\pm\sim0.2\%$)
near-sinusoidal photometric ($V$-band)
variability in GD\,356, with a period of
0.0803~days ($\sim115$~minutes). We interpret this as the rotation
period of the star. We model the variability with a dark spot (by
analogy with star spots) covering $10\%$ of the stellar surface. It
seems likely that this spot is also the site of the Zeeman emission,
requiring the presence of a temperature inversion. 
We show that the spot is never totally visible or obscured, and that
both polar and equatorial spots
produce good fits to the data when viewed at high and low inclination
respectively.

\end{abstract}


\section{Introduction}

Isolated magnetic white dwarfs are generally recognised by their
Zeeman lines of hydrogen or helium in absorption. Studies of local
samples of white dwarfs suggest $\sim10\%$ or more
are magnetic in the range $10^4-10^9$G (Liebert, Bergeron \& Holberg
2003), and Schmidt \etal (2003) have recently doubled the number of
catalogued magnetic white dwarfs to $\sim120$ from discoveries in the
Sloan
Digitised Sky Survey. Many magnetic white dwarfs display surface
inhomogeneities, which may be linked to field structure or interstellar
accretion, from which their rotation periods can be derived
(e.g. RE\,J$0317-853$, $P_{\rm rot}\sim12$~minutes, Barstow \etal
1995). It is notoriously difficult to determine such information for
white dwarfs in general, since their absorption lines are
significantly broadened by the high surface gravity. Thus, studies of
individual magnetic white dwarfs are important for understanding the
evolution of the population in general, and the origin of the magnetic
sample in particular.

GD\,356 (B $=13$\,MG)
is unique among magnetic white dwarfs in showing resolved
Zeeman triplets of H$_\alpha$ and H$_\beta$ in emission (Greenstein \&
McCarthy 1985). Detailed modelling of spectropolarimetric observations
by Ferrario \etal (1997) points to the existence of a latitudinally
extended spherical sector or strip covering approximately 10\% of the
stellar surface, over which the stellar atmosphere has an inverted
temperature distribution at low optical depths. This small region
is most likely the site of origin of the emission lines.

The cause of this temperature inversion is a mystery, especially as there
is
no evidence for a low-mass close, stellar companion. Ferrario \etal (1997)
investigated the most obvious interpretation, chromospheric activity,
but the evidence for a very localised emission region, and the absence
of evidence for such a phenomenon in other similar magnetic white
dwarfs argues against such an interpretation. Ferrario \etal (1997)
also considered a model invoking Bondi-Hoyle accretion from the
interstellar medium, but this interpretation remains unsatisfactory
due to a lack of detectable x-ray emission.

Other authors have considered more exotic explanations for the
presence of Zeeman-split emission lines in GD\,356. For example,
Li, Ferrario \&
Wickramasinghe (1998) suggest that an Earth-like planet, orbiting
through the magnetic field with a period of a few hours, might heat the
white dwarf's atmosphere near the poles via the generation of
electrical currents. Alternatively, and following the work of
Zheleznyakov and Serber (1995), Gnedin \etal (2001) propose that
in the regions of the magnetic poles cyclotron radiation pressure
exceeds the local force of gravity, driving a plasma jet into the
magnetosphere. The emission lines may be formed in this outflowing
plasma.

\begin{figure*}

\psfig{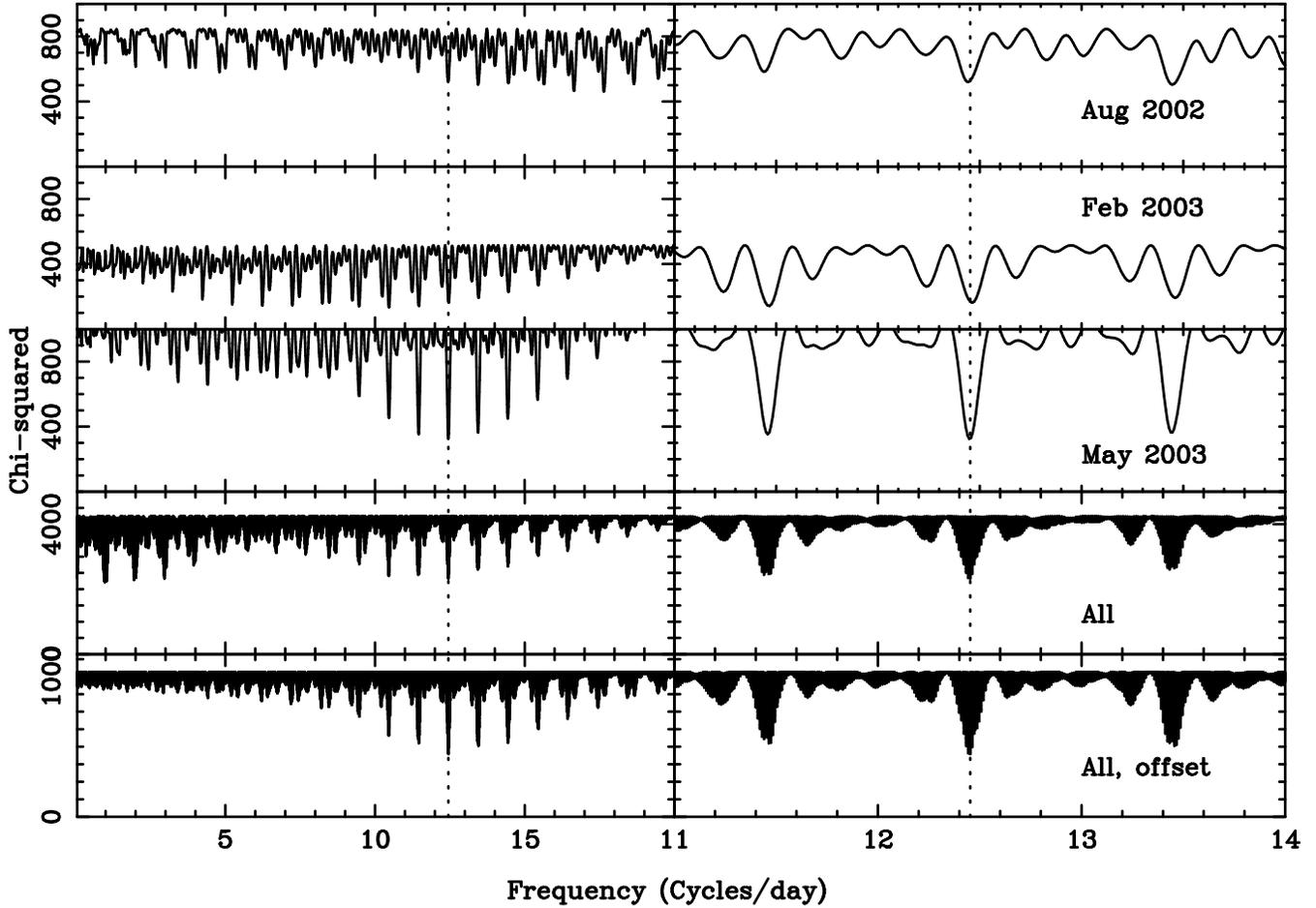}

\caption{Periodograms for all of the data sets. Panel 1: August
  2002. Panel 2: Feb 2003. Panel 3: May 2003. Panel 4: all data.
  Panel 5: all data, with the average y value of each set offset to
  zero, removing slight variations in the differential magnitude
  between each epoch (see Section 3). A period of 0.080300 days is
  favoured (dotted line).} 
\label{fig1}
\end{figure*}

Crucially, none of these models were able to utilise information about
the rotation period of GD\,356, since no evidence for rotation had
previously been detected. Ferrario et al.'s multi-epoch spectropolarimetry
appeared to rule out rotation in the period range $\sim 1$~h -- 3~year,
and a lack of variability in Gnedin et al.'s long-term spectropolarimetric
observations led them to conclude that the rotation period of GD\,356
exceeds 5~years. Indeed, Ferrario et~al. (1997) admit that the details of
the star's underlying field geometry are difficult to ascertain with any
certainty given the data available to them, especially a lack of
rotational modulation.

Here, we present precise $V$-band photometric observations of GD\,356
which reveal low amplitude ($\pm \sim 0.2\%$) near-sinusoidal variability
with a period of 0.0803~days ($\approx 115$~minutes). We argue that this
variability is due to rotation, consistent with the presence of a cool,
dark spot (analagous to a sunspot) covering a small fraction of the
stellar surface. Overlying this region is the temperature inversion at low
optical depths, as proposed by Ferrario et al. (1997), which is the site
of origin of the emission lines.

\begin{figure}

\psfig{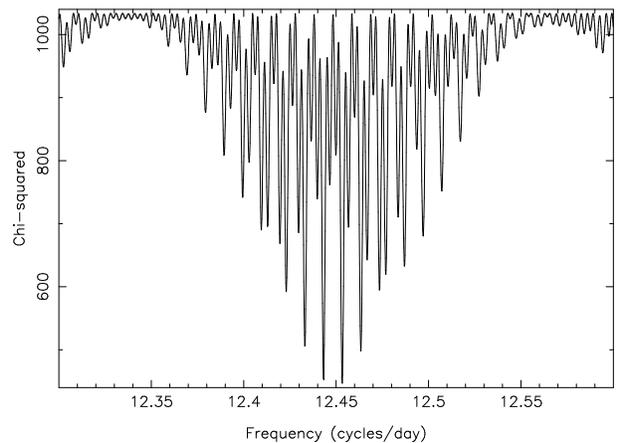}

\caption{Periodogram zoomed in on the best fitting period of 12.4533 cycles/day.} 
\label{fig2}
\end{figure}

\section{Observations}

We observed GD356 at 3 epochs, in August 2002, February 2003 and May
2003 using the 1.0-m Jacobus Kapteyn telescope on La Palma. We used
the SITe1 CCD chip, which is 2088x2120 pixels, with readout noise = 6 e/ADU
and gain = 1.9. The
observtions were all taken using a V Harris filter with a typical exposure time
of 60-90s. The chip was windowed to improve the readout time for for the
August and May runs, but not for the February run due to reduction
software problems at the time. See Table~1 for a full list of
observations.

\begin{table}
\caption{List of observations of GD356 taken with the Jacobus Kapteyn
  Telescope on La Palma.}
 \begin{tabular}{cccccc}
Date & Exp & N & Conditions \\  
dd/mm/yyyy & (s) & &  \\ 
& & & \\   
01/08/02  & 60 & 45 & Clear, seeing 1.5''\\ 
03/08/02  & 60 & 10 & Clear, seeing 0.7-1''\\
04/08/02  & 60 & 10 & Clear, seeing 0.8-1.3''\\
05/08/02  & 90 & 15 & Variable. Seeing 0.7-1.5''\\
06/08/02  & 60 & 55 & Clear, seeing 0.7''\\
07/08/02  & 60 & 31 & Clear, seeing 0.8''\\
21/02/03  & 120 & 34 & Clear, seeing 1.4-2''\\
22/02/03  & 90 & 20 & Cirrus, seeing 2''\\
25/02/03  & 60 & 27 & Clear, seeing 1-1.2''\\
26/02/03  & 90 & 28 & Clear, seeing 1.5''\\
25/05/03  & 85 & 66 & Clear, seeing 1''\\
27/05/03  & 120 & 50 & Clear, seeing 1''\\
30/05/03  & 120 & 40 & Clear, seeing 1''\\
31/05/03  & 120 & 50 & Clear, seeing 0.8''\\
\end{tabular}
\end{table}

\section{Data reduction}

%CSB, include reduction and derivation of period

The data were all reduced using TRM's ULTRACAM pipeline software. 
Bias frames for each night were combined to create a master bias,
which was then subtracted from all other frames. Sky flats were then
checked, and any with counts of less than 10000 or greater than 35000
were discarded. The remaining flats were combined to create a master
flat for each night that was normalised and divided from the target 
frames. There were no sky flats for the 22nd, 25th or 26th of February so 
the target frames from those nights were flat-fielded using combined 
dome flats.

Differential photometry was performed on the target with respect
to 2 bright comparison stars in the field. Once we had established
that neither was varying, we then used the data that was reduced with
respect to the brightest comparison. Results were output in
differential magnitudes.

We used a ``floating mean'' periodogram ( e.g. Cumming, Marcy \&
Butler 1999,
Morales-Rueda \etal 2003) to determine the period of  each epoch separately,
and all of the data together. This is a generalisation of the
Lomb-Scargle periodogram (Lomb 1976, Scargle 1982) and involves
fitting the data with a sinusoid plus constant of the form:

\[
 A + B \sin[2\pi f(t - t_{0})],
\]

\noindent
where $f$ is the frequency and $t$ is the observation time. The
advantage over the Lomb-Scargle periodogram is that it treats the
constant, A, as an extra free parameter rather than fixing the
zero-point and then fitting a sinusoid, i.e. it allows the zero-point
to ``float'' during the fit. The resultant periodogram is an inverted
$\chi^2$ plot of the fit at each frequency. Periodograms of each
epoch and all of the data combined can be seen in Fig 1. 

There is a small offset in differential magnitude between each epoch 
which may be a sign of a second, longer-term, variability, but more 
data will be required to verify this result. In order to obtain a
clearer periodogram for the short-term variation, we removed this
long-term variation by offsetting the average y value of each data set 
to zero, hence the August, February and May data sets were offset in y by 
-0.208, -0.206 and -0.204 differential mags respectively. The bottom panel 
of Fig~1 shows the periodogram for all the data when the long-term variation
has been removed. As the standard error estimates output by the ULTRACAM 
software are
likely to be an underestimate due to the relatively high
signal-to-noise of the data, the error bars for this last data set 
were re-scaled by adding
0.001 in quadrature to the quoted errors to bring the reduced
chi-squared down to 1. This data set was then used to generate the
best-fit period and ephemeris.

Two periods were found to be significantly (difference in chi-squared
$> 60$) better than the rest (see Fig 1): 

\[
HJD = 2452715.07580(5) + 0.0803000(7)E 
\]
and 
\[
HJD = 2452715.97555(5) + 0.0803652(3)E
\]
where the ephemeris given is the point of minimum light for which the
correlation between the fitted zero point and period is at a minimum. 

The shorter period is marginally favoured as its chi-squared is 4
less than that of the longer period.

\section{A cool spot variability model}

The analysis of spectropolarimetric observations of GD\,356 undertaken by
Ferrario et al. (1997) led those authors to conclude that magnetic
activity is confined to a limited, latitudinally extended spherical sector
or strip which covers around 10\% of the stellar surface. The presence of
H$_\alpha$ and H$_\beta$ in emission shows that a temperature inversion
exists in this region, but since the gas is tenuous and optically thin it
is energetically insignificant and unlikely to contribute to the continuum
emission. By analogy with star- and sun-spots, where convective energy
flow is supressed by high magnetic fields (at $T_{eff}\sim7500$\,K,  
GD\,356 has a convective atmosphere, Bergeron, Leggett \& Ruiz 2001), 
we expect regions of higher
field such as this to be dark. In this section we aim to show that the
near sinusoidal variability visible in our $V$-band light cure is
consistent with this dark spot.

\begin{figure*}

\psfig{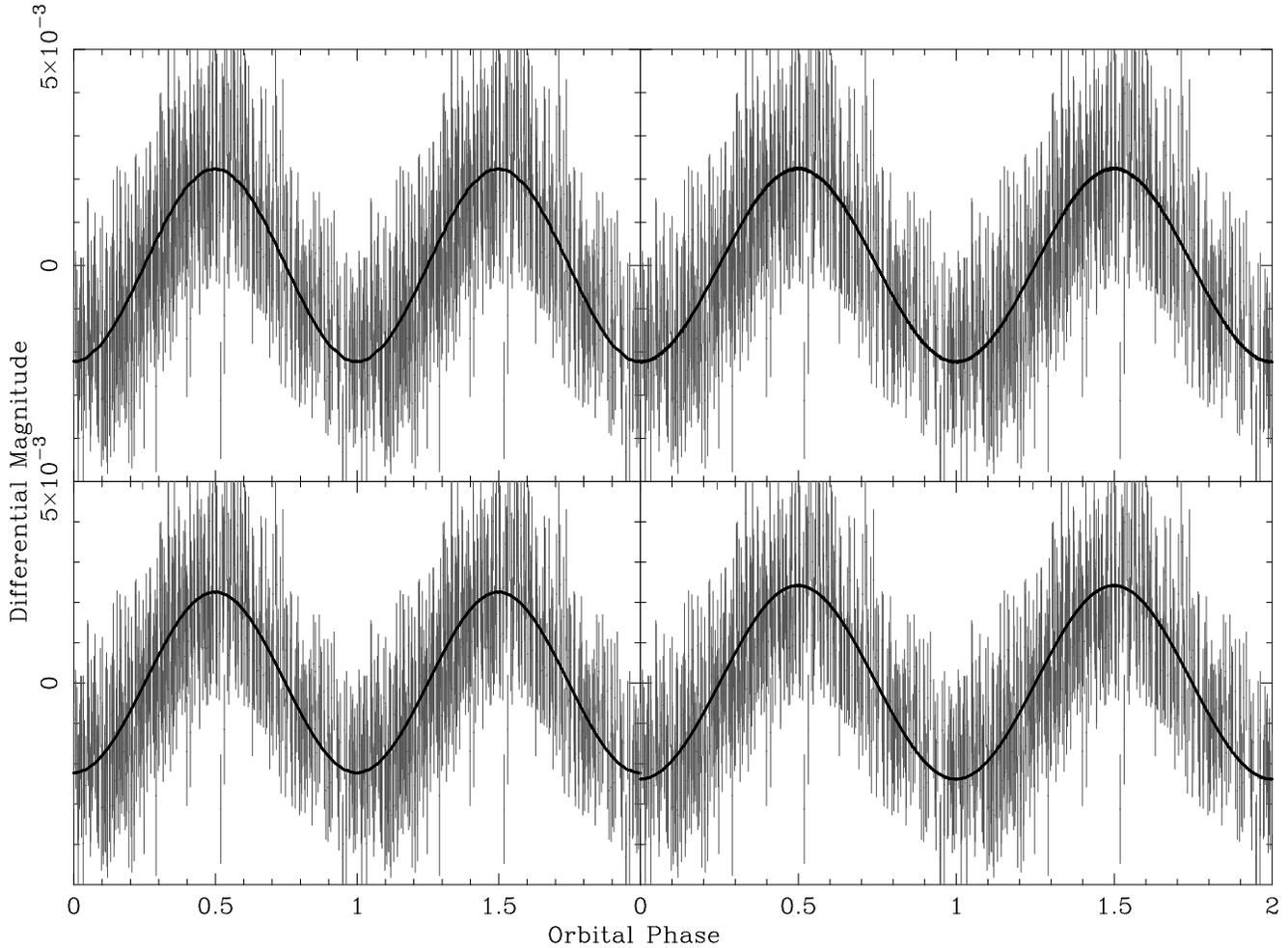}

\caption{Phase-folded light curve for all data with the slight
variations in differential magnitude between each epoch removed (see
Section 3). Data is folded on a period of 0.080300 days and
  fit with 4 different models: top left: $i=0.7^\circ$, $\theta_m =
  60^\circ$, $L_s = 0.1$; bottom left: $i=1.4^\circ$, $\theta_m =
  60^\circ$, $L_s = 0.5$; top right:  $i=60^\circ$, $\theta_m =
  0.7^\circ$, $L_s = 0.1$; bottom right: $i=60^\circ$, $\theta_m =
  1.5^\circ$, $L_s = 0.5$. See Section 4 for further details. Vertical bars are errors in differential magnitude, after correction to give a reduced chi-squared of 1 in the periodogram (see Section 3).} 
\label{fig3}
\end{figure*}

We assume this spot to be a circular area of the
stellar surface which subtends an angle $\beta$ around colatitude 
$\theta_m$, as measured from the rotation axis of the white dwarf.
King and Shaviv (1984) discussed the form
of the light curves produced by self occultation of a hot emission
region by the white dwarf body. A dark spot can produce quasi-sinusoidal light curves in a similar
manner when the spot is never totally visible or obscured. The following conditions on the positions of maximum visibility
\begin{equation}
| i - \theta_m | + \beta  >  90^\circ
\label{eq:cond1}
\end{equation}
and, minimum visibility  
\begin{equation}
i + \theta_m + \beta  >  90^\circ
\label{eq:cond2}
\end{equation}
of the emission region must be satisfied, where $i$ is the inclination 
of the observer's line of sight to the stellar rotation axis. For a
given $\beta$ the light curve is a function of the position of the 
system in the plane $0^\circ \le i \le 90^\circ$, $0^\circ \le \theta_m \le 90^\circ$.
Following the work of Ferrario \etal 
we fix $\beta = 40^\circ$ giving the region a fractional
area of $\sim 0.1$. The small amplitude of the variations in the
observed light curve of GD\,356  
($\pm\sim 0.2\%$) leads to a further constraint on the angles. To produce
such small amplitude variations the angular difference between the
positions of maximum and minimum visibility ($\delta$) must be small,
leading to the constraint
\begin{equation}
| i - \theta_m | - (i + \theta_m) = \delta \rightarrow 0^\circ.
\label{eq:cond3}
\end{equation}
Here we have implicitly assumed that the spot is much darker than the surrounding stellar
atmosphere. As discussed below this assumption is good if 
the spot has a relative luminosity compared to the mean
global luminosity of the WD $L_s \le0.5$. 
Combining conditions (1), (2) and  
(3), and using $\beta = 40^\circ$ we find that the
inclination angle and the colatitude of the cool region 
must obey either $(\theta_m > 50^\circ, i \rightarrow 0^\circ)$ or 
$(i > 50^\circ, \theta_m \rightarrow 0^\circ)$. Hence, within the 
assumptions outlined above,  we are either observing the rotational 
modulation of a dark region near the rotational 
equator from a position near the rotational pole, or 
conversly, we are observing a spot close to the rotational pole
from high inclination.

To investigate the form of the light curve we calculated the visible
fraction of the dark spot over a rotational cycle, based on a 
simple model similar to Wynn and King (1992). 
The spot was divided into a number ($> 100$) of equal area 
sectors with a constant luminosity relative to the surrounding stellar 
surface. Figure 3 contrasts the data and the model light curves.
Both polar and equatorial spots produce a good fit to the data when 
viewed at high and low inclination respectively. The 
relative luminosity of the spot makes little difference to the values
of $i$ and $\theta_m$ required to fit the data for $L_s \le 0.5$. 
Of course, there is a large change in the angles required to fit the
data as $L_s  \rightarrow 1$. 

\section{Discussion}

The low amplitude ($\pm\sim0.2\%$), 
0.0803~day near-sinusoidal photometric ($V$-band) variability of
GD\,356 is consistent with the rotational modulation 
of a dark spot covering ($\sim 10\%$), and which is never totally
visible or obscured. This spot
is either observed near the rotational equator from a position near
the rotational pole, or close to the rotational pole from high
inclination. The region is presumably the source of the Zeeman-split
emission lines of H$_\alpha$ and H$_\beta$ seen in optical spectra, as
proposed previously by Ferrario \etal (1997).
Although those authors (and Gnedin
\etal 2001) did not report variability of the emission lines, either
in velocity or flux, in the light of the discovery of broad-band
photometric variability a detection of modulation of the emission lines 
themselves may provide clues as to their origin. However, this
photometric modulation is so small that, even if it is stronger in the 
Zeeman components than the
photometry, it may be difficult to detect. We note that while we have concentrated on the dark-spot model as the source of the photometric variability, we are unable to rule out a grey-spot model (ie a spot which is not cool enough to appear dark, extending over a larger area), or a non-uniform temperature distribution across the surface of the white dwarf, unrelated to the chromospheric spot. 

We have deliberately not attempted here to give a physical explanation
for the origin of the Zeeman-split emission lines that make GD\,356 unique
even among the rare magnetic white dwarfs.  Chromospheric activity (and an
associated x-ray bright corona) is an obvious explanation, but no evidence
for such a phenomenon has ever been detected from a magnetic white dwarf
(e.g.  Arnaud, Zheleznyakov and Trimble 1992, Cavallo, Arnaud and Trimble
1993). From {\em ROSAT} observations, Muielak, Porter and Davis (1995)
give the upper limit on the x-ray luminosity of GD\,356 at a few $\times
10^{26}$~erg s$^{-1}$, the same order of magnitude as the luminosity of
the solar corona. The only white dwarf that might have a corona detected
at x-ray wavelengths, KPD\,0005$+$5106, has an implied luminosity five
orders of magnitude higher (Fleming, Werner and Barstow 1993).  The
evidence for a very localised emission region led Ferrario et al. (1997)
to also argue against the chromospheric activity / hot corona
interpretation.

Alternatively, GD\,356 might be accreting either from a nearby companion
or via Bondi-Hoyle accretion from an interstellar medium.  However,
near-infrared photometry presented by Ferrario et al. (1997) rules out the
presence all but very low mass non-stellar companions, and the lack of
detectable
x-ray emission by
Musielak, Porter and Davis (1995) renders the accretion scenario
unsatisfactory at present.

We note, though, that new, more sensitive x-ray observations of GD\,356
have recently been scheduled with the {\em Chandra} observatory. The
detection of hard x-ray emission, and possibly x-ray variability on the
same timescale as the optical oscillations reported here would give
credibility to the chromospheric activity and/or accretion models.

\section{Acknowledgements}

CSB acknowledges the support of a PPARC studentship. MRB, GAW and TRM
also acknowledge the support of PPARC. We thank the anonymous referee for helpful comments. 

 The Jacobus Kapteyn Telescope 
is operated on the island of La Palma
by the Isaac Newton Group in the Spanish Observatorio del
Roque de los Muchachos of the Instituto de
Astrofisica de Canarias. This research has made use of the SIMBAD
database, operated at CDS, Strasbourg, France.


\begin{thebibliography}{99}

\bibitem{}
Arnaud K.A., Zheleznyakov V.V., Trimble V., 1992, PASP, 104, 239

\bibitem{}
Barstow M.A., Jordan S., O'Donoghue D., Burleigh M.R., Napiwotzki R.,
Harrop-Allin M.K., 1995, MNRAS, 277, 971

\bibitem{}
Bergeron P., Leggett S.K., Ruiz M.T., 2001, ApJS, 133, 413

\bibitem{}
Cavallo R., Arnaud K.A., Trimble V., 1993, JApA, 14, 141

\bibitem{}
Cumming A., Marcy G.W., Butler R.P., 1999, ApJ, 526, 890

\bibitem{}
Ferrario L., Wickramasinghe D.T., Liebert J., Schmidt G.D., Bieging
J.H., 1997, MNRAS, 289, 105

\bibitem{}
Fleming T.A., Werner K., Barstow M.A., 1993, ApJ, 416, L79

\bibitem{}
Gnedin Y. N., Borisov N.V., Natsvlishvili T.M., Piotrovich M.Y., 2001,
Astrophysics, 44, 321

\bibitem{}
Greenstein J.L., McCarthy J.K., 1985, ApJ, 289, 732

\bibitem{}
King A.R., Shaviv G., 1984, MNRAS, 211, 883

\bibitem{}
Li J., Ferrario L., Wickramasinghe D., 1998, ApJ, 503, L151

\bibitem{}
Liebert J., Bergeron P., Holberg J.B., 2003, AJ, 125, 348

\bibitem{}
Lomb N.R., 1976, Ap\&SS, 39, 447

\bibitem{}
Morales-Rueda L., Maxted P.F.L., Marsh T.R., North R.C., Heber U.,
2003, MNRAS, 338, 752

\bibitem{}
Musielak Z.E., Porter J.G., Davis J.M., 1995, ApJ, 453, L33

\bibitem{}
Scargle J.D., 1982, ApJ, 263, 835

\bibitem{}
Schmidt G.D., et al., 2003, ApJ, in press

\bibitem{}
Wynn G.A., King A.R., 1992, MNRAS, 255, 83

\bibitem{}
Zheleznyakov V.V., Serber A.V., 1995, Adv. Space Res., 16, 3, 77

\end{thebibliography}
\end{document}